\documentstyle[12pt,aps,epsf,floats]{revtex}
\def\kt{$k_T$}
\def\mkt{k_T}
\def\avkt{$\langle k_T \rangle$}
\def\mavkt{\langle k_T \rangle}
\def\avktsq{$\langle k_T^2 \rangle$}
\def\mavktsq{\langle k_T^2 \rangle}
\def\pt{$p_T$}
\def\mpt{p_T}
\def\avpt{$\langle p_T \rangle$}
\def\avptp{$\langle p_T \rangle_{pair}$}

\def\sp{$\sigma_{1parton,2D}$}
\def\msp{\sigma_{1parton,2D}}

\def\msg{\sigma_{\gamma,1D}}
\def\sgtwo{$\sigma_{\gamma,2D}$}
\def\msgtwo{\sigma_{\gamma,2D}}
\def\x{$x$}
\def\s{$\sqrt{s}$}
\def\DZERO{D\O}

\begin{document}

\title{
\hspace*{\fill}\parbox[b]{3.4cm}{\small 
CTEQ-805 \\
MSUHEP-80501 \\  
UR-1539 \\
\today \\
}	
\\
$k_T$ Effects in Direct-Photon Production}
\author{L. Apanasevich, C. Bal\'azs, C. Bromberg, J. Huston, A. Maul,
W. K. Tung}
\address{Department of Physics and Astronomy, Michigan State University,
East Lansing, MI 48824, USA}
\author{S. Kuhlmann}
\address{
High Energy Physics Divison, Argonne National Laboratory, Argonne, IL 60439, USA}
\author{J. Owens}
\address{High Energy Physics, Florida State University, Tallahassee, FL 32306, USA}
\author{M. Begel, T. Ferbel, G. Ginther, P. Slattery, M. Zieli\'nski}
\address{Department of Physics and Astronomy, University of Rochester,
Rochester, NY 14627, USA}
\maketitle

\begin{abstract}

We discuss the phenomenology of initial-state parton-\kt\ broadening in 
direct-photon production and related processes in hadron collisions.
After a brief summary of the theoretical basis for a Gaussian-smearing 
approach, we present a systematic study  of recent results on fixed-target 
and collider direct-photon production, using complementary data 
on diphoton and pion production to provide empirical guidance on the 
required amount of \kt\ broadening.  This approach provides a consistent 
description of the observed pattern of deviation of next-to-leading order 
QCD calculations relative to the  direct-photon data, and accounts for 
the shape and normalization difference between fixed-order perturbative 
calculations and the data. We also  discuss the uncertainties in this 
phenomenological approach, the implications of these results on the 
extraction of the gluon distribution of the nucleon, and the comparison 
of our findings to recent related work.
\end{abstract}

\section*{Introduction}

	Direct-photon production has long been viewed as an ideal 
vehicle for measuring the gluon distribution in the 
proton \cite{halzen}. The quark-gluon Compton scattering subprocess 
($gq{\rightarrow}{\gamma}q$) 
dominates $\gamma$ production in all kinematic regions of $pp$ scattering, 
as well as
for low to moderate values of parton-momentum fraction, \x, in
$\overline{p}p$ scattering; the cross sections have been calculated 
to next-to-leading order (NLO) \cite{aurenche}.
The gluon distribution in the proton is relatively
well-constrained at small \x\ ($x<$0.1) by deep-inelastic scattering (DIS) and 
Drell-Yan (DY) data, but less so at larger~\x\ \cite{cteqgluon}.
Consequently, direct-photon data from fixed-target 
experiments that have been incorporated in several modern global parton 
distribution function analyses can, in principle, provide a major
constraint on the gluon content at moderate to large \x\ 
\cite{mrs,cteq4,abfow,grv}.

	However, a pattern of deviations between the measured 
direct-photon cross sections and NLO calculations has been 
observed \cite{cteqkt}. The discrepancy is particularly
striking in the recently published higher-statistics
data from E706 experiment \cite{e706}.
E706 observed large deviations between NLO calculations 
and data, for both direct-photon and ${\pi}^{0}$ inclusive cross sections. 
The final direct-photon results from UA6 \cite{ua6} 
also exhibit evidence of similar, although smaller, discrepancies.
The suspected origin of the disagreements is 
from effects of initial-state soft-gluon radiation.
Such radiation generates transverse components 
of initial-state parton momenta, referred to in this discussion as
${k}_{T}$. To be precise, as in \cite{e706},
\kt\ denotes the magnitude of the effective transverse momentum vector, 
${\vec k}_T$, of each of the two colliding partons.

\begin{figure}[t]
\begin{center}
\epsfxsize=9cm
\epsfysize=12cm
\vskip-1.cm
\mbox{\epsfbox{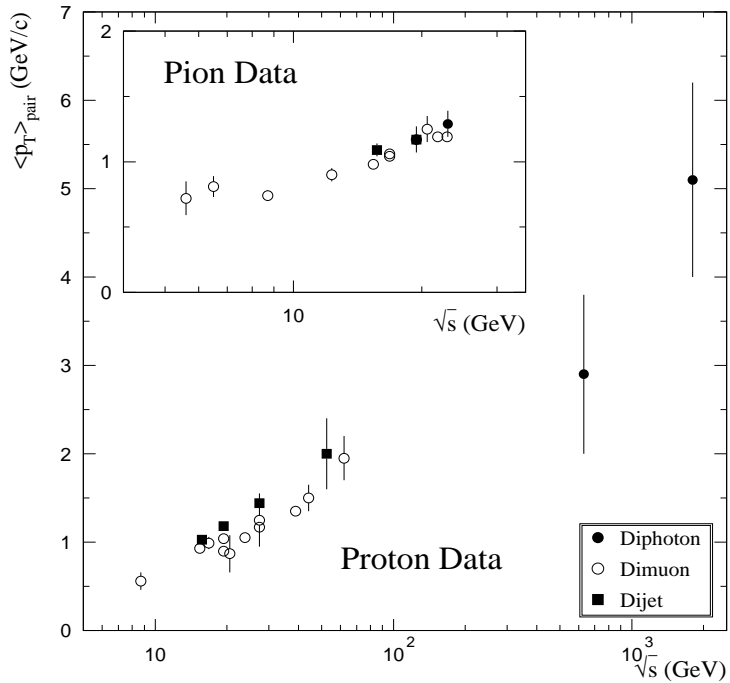}}
\vskip-1.cm
\end{center}
\caption{
\avpt\ of pairs of muons, photons, and jets produced in
hadronic collisions versus $\sqrt{s}$. 
} 
\label{fig:pairpt}
\end{figure}

Evidence of significant \kt\ has long been observed in
measurements of dimuon, diphoton, and dijet pairs.
A collection of measurements of the average 
transverse momentum of the pairs
(\avptp) is presented in Fig.~\ref{fig:pairpt}, for a wide range of 
center-of-mass energies ($\sqrt{s}$) \cite{pairpt}.

The values of \avptp\ are large, and  they increase
approximately logarithmically with increasing $\sqrt{s}$.
The values of \avkt\ per parton (estimated as $\approx$\avptp/$\sqrt{2}$)
indicated by these DY, diphoton, and dijet
data, as well as the inclusive direct-photon and $\pi^0$ production data,
are too large to be interpreted as ``intrinsic" --- i.e., due only to the 
finite size of the proton. 
(From these data, one can infer that the average ${k}_{T}$ per parton
is about 1 GeV/$c$ at fixed-target energies, increasing 
to 3--4 GeV/$c$ at the Tevatron collider.
One would expect \avkt\ values on the order of
0.3--0.5 GeV/$c$ based solely on proton size.)
Perturbative QCD (PQCD) corrections at NLO level are also  insufficient
to explain the size of the observed effects, and full resummation calculations
are required to describe DY and W/Z \cite{ellis,WZ,resbos}, and diphoton 
\cite{fergani,diphoton} distributions.
Values of \avptp\ for DY and diphoton data exhibit similar trends
versus energy; for DY data, pion-beam values are somewhat larger than 
those for proton beams at the same $\sqrt{s}$. 
The dijet data hint at somewhat larger values of \avptp\
than DY results at the same energy, a difference that may be related to different
color-flow between initial and final states in DY and in dijet
events, as well as to  a larger contribution from gluon-induced subprocesses 
for dijet production.
Similar soft-gluon (or \kt) effects are expected to be present
in all hard-scattering processes, such as the inclusive production of jets 
or direct photons \cite{FF}. 

This paper presents a phenomenological model for $k_T$ effects in 
direct-photon production and, by extension, in all hard-scattering 
processes. Quantitative comparisons of this model with data from
E706 have been reported previously \cite{e706,altarelli}.
We will discuss the successes and uncertainties of this prescription, as well as 
the implications for determining the gluon distribution. 

\section*{\kt\ Smearing Formalism: Theory and Practice}

We now briefly describe the theoretical underpinnings of \kt\ effects
using the Collins-Soper-Sterman (CSS) formalism \cite{CSS}.
In this formalism, the \pt\ spectrum in hard-scattering 
processes is written as the convolution of the parton distributions $f_{i/h}$, 
the $C$ functions (representing the finite pieces of the virtual corrections), 
and two Sudakov form factors, ${\cal S}^P$ and ${\cal S}^{NP}$.
${\cal S}^P$ can be regarded as being perturbative in character and 
${\cal S}^{NP}$ as non-perturbative. The perturbative Sudakov form factor 
represents a formal resummation of soft-gluon emissions. The non-perturbative 
Sudakov form factor is determined from a fit to the data, is expected to be 
universal (for a given parton flavor and hadron type), 
and is usually parameterized as a Gaussian distribution. 
The dividing line between the perturbative and non-perturbative 
contributions is somewhat arbitrary (similar to the better known cases of parton 
distributions and fragmentation functions), and it is quantified by a 
theoretical scale in the resummation formalism. The ${p}_{T}$ distribution
(e.g., of the Drell-Yan pair) can be written symbolically as
% EEEEEEEEEEEEEEEEEEEEEEEEEEEEEEEEEEEEEEEEEEEEEEEEEEEEEEEEEEEEEEEEEEEEEE
  \begin{eqnarray}
  {d \sigma \over dp_T} =
  \sigma_0 ~ e^{-({\cal S}^{NP}+{\cal S}^{P})} \otimes 
           \left[ 
              \left( C_{a_1 i} \otimes f_{i/h_1} \right)
              \left( C_{a_2 j} \otimes f_{j/h_2} \right)
           \right]
      + Y.
  \label{}
  \end{eqnarray}
% eeeeeeeeeeeeeeeeeeeeeeeeeeeeeeeeeeeeeeeeeeeeeeeeeeeeeeeeeeeeeeeeeeeeeee
The $Y$ function is added to ensure a smooth matching between the low and 
the high-$p_T$ regions, where the resummed and the fixed-order descriptions
work better, respectively.

	At collider energies, most of the $k_T$ can be attributed to
perturbative soft-gluon emission. However, for fixed-target kinematics, 
almost all of the $k_T$ is due
to the non-perturbative mechanisms. A proper treatment requires both the 
appropriate data to determine the non-perturbative input, and an implementation
of the soft-gluon resummation formalism for the particular process.

	The resummation calculation for multiple soft-gluon emission 
in direct-photon production is quite
challenging. The production rate and kinematic distributions of photon
pairs produced in hadron interactions have already been calculated 
\cite{diphoton}, and a similar calculation of the transverse momentum 
distribution of a photon-jet system is also plausible, but more involved, 
since the final-state parton takes part in soft-gluon emission 
and in color exchange 
with the initial-state partons. A recent work on this subject \cite{lai}
addressed only the effects of multiple soft-gluon radiation 
in the initial state. Incorporating jet definition in the formalism 
is also not a fully resolved issue. Finally, the calculation of individual 
transverse momenta of the photon and the jet is further complicated by the
fact that several overlapping power-suppressed corrections can contribute.
In the absence of the full resummed calculation, approximations are made
in order to compare theory with data.

	In lieu of a rigorous calculation of the resummed $p_T$
distribution, effects of soft-gluon radiation can be approximated  by a 
convolution of the leading-order cross section with a ${k}_{T}$-smearing 
function \cite{owensll}. In the formalism described above, this is equivalent
to absorbing all of the perturbative gluon emissions into the non-perturbative
Sudakov form factor. Since no explicit resummation of soft-gluon emissions is 
performed, the average value of ${k}_{T}$ used in the 
smearing should be representative of the value observed (or expected) 
in the kinematic regime of the experiment.

	The expression for the  leading-order (LO) cross section 
for direct-photon production at large \pt\ has the form:
\begin{eqnarray}
&& E_\gamma \frac{d^3\sigma}{dp^3}(h_1h_2\to \gamma X) = 
\nonumber \\ &&
\sum_{a_1a_2a_3} \int dx_1 dx_2 \; f_{a_1/h_1}(x_1,Q^2) \; f_{a_2/h_2}(x_2,Q^2) 
\frac{\hat{s}}{\pi} \frac{d\sigma}{d\hat{t}}(a_1a_2\to \gamma a_3) 
\; \delta(\hat{s}+\hat{t}+\hat{u}),
\end{eqnarray}
where $d\sigma/d\hat{t}$ is the hard-scattering matrix element,
and $f_{a_1/h_1}$ and $f_{a_2/h_2}$ are the parton distribution functions 
(pdf) for the colliding partons $a_1$ and $a_2$ in hadrons $h_1$ and $h_2$, 
respectively. To introduce $k_T$ degrees of freedom, one extends each integral 
over the parton distribution functions to the ${k}_{T}$-space, 
\begin{equation}
d{x}_{1} \; {f}_{a_1/h_1}({x}_{1},{Q}^2)~{\rightarrow}~d{x}_{1}{d}^2{k}_{T_1}
\; g({\vec k}_{T_1}) \; {f}_{a_1/h_1}({x}_{1},{Q}^2),
\end{equation}
(a corresponding substitution is done for parton $a_2$ in hadron $h_2$).

The distribution $g({{\vec k}_{T}}$) is usually taken to be a Gaussian,
\begin{equation}
g({{\vec k}_{T}})={{e^{-{k}_{T}^2/\mavktsq}}\over{{\pi}\mavktsq}},
\end{equation}
where \avktsq\ is the square of the 
2-dimensional (2D) RMS width of the ${k}_{T}$ distribution for one parton 
($\sigma^2_{1parton,2D}$), and
is related to the square of the 2D average of the absolute value of
${\vec k}_T$ of one parton through \avktsq\ = 4$\mavkt^2/\pi$. 
We emphasize that \avkt\ represents the average effective 2D transverse momentum
per parton. (The average transverse momentum of the parton pair is, of course,
a factor of $\sqrt{2}$ larger than the average transverse momentum per parton.)

The 4-vectors of the colliding partons are expressed in terms of the 
momentum fraction \x\ of the partons. Ignoring parton and hadron masses,
\begin{equation}
{x}_{1}=({E}_{1}+{p}_{l_1})/\sqrt{s},
\end{equation}
where the parton four-vector is
\begin{equation}
{p}_{1}=({E}_{1},{{\vec k}_{T_1}},{p}_{l_1}),
\end{equation}
with
\begin{equation}
{E}_{1}={1\over 2} \left[{x}_{1}{\sqrt s} + {{k}^2_{T_1}\over {x}_{1}{\sqrt{s}}}\right],
\end{equation}
and
\begin{equation}
{p}_{l_1}={1\over 2}\left[{x}_{1}{\sqrt s} - {{k}^2_{T_1}\over{x}_{1}{\sqrt{s}}}\right].
\end{equation}
(Similar expressions are used for parton $a_2$.)

It is straightforward to evaluate the invariant cross sections, 
including ${k}_{T}$ effects, according to  the above prescription.
In general, because the unmodified PQCD cross sections fall rapidly 
with increasing \pt, the net effect of the \kt\ smearing is to increase 
the yield. We denote the enhancement factor as $K(\mpt)$.
Since the invariant cross section for direct-photon 
production is now a six-dimensional integral, it is convenient to employ 
Monte Carlo techniques in the evaluation of $K(\mpt)$.
An exact treatment of the kinematics can be implemented in a Monte 
Carlo framework, but it is more difficult in an analytic approach.

	A Monte Carlo program that includes such a treatment of
${k}_{T}$ smearing, and the leading-order cross section
for high-\pt\ particle production, has long been available \cite{owensll}.
The program provides calculations of many experimental observables,
in addition to inclusive cross sections. The program can be used for 
direct photons, jets, and for single high-\pt\ particles resulting from 
jet fragmentation (such as inclusive $\pi^0$ production).
Unfortunately, no such program is available for NLO calculations, but
one can approximate the effect of \kt\ smearing by multiplying the NLO 
cross sections by the LO  ${k}_{T}$-enhancement factor.
Admittedly, this procedure involves a
risk of double-counting since some of the ${k}_{T}$-enhancement may 
already be contained in the NLO calculation. However, such double-counting
effects are expected to be small, and consequently this uncertainty on 
$K(\mpt)$ should also be small.  (For example, the NLO estimate
for \avptp\ of the diphoton pairs produced in $pp\rightarrow\gamma\gamma$ 
at $\sqrt{s}$ = 31.5 GeV is on the order of a few hundred MeV/$c$, while 
the resummed prediction is well over 1 GeV/$c$ \cite{resbos}.)

It is clear that this type of treatment of $k_T$ effects is model dependent.
In particular, different functional forms can lead to quantitatively 
different answers. For example, adding substantial
non-Gaussian tails in \kt\ smearing can affect the output distributions.
One of the strengths of the approach we follow in this paper
is that the ${k}_{T}$ distribution used in the smearing is based
on experimental information, and the Gaussian character of the ${k}_{T}$ 
effects is consistent with the data observed by E706 \cite{e706}.
Moreover, any non-Gaussian tails may result primarily from single  
hard-gluon emission, and such contributions should therefore 
already be contained in the NLO cross sections used in this analysis. 

A complete treatment of soft-gluon radiation in high-\pt\
production, including the appropriate non-perturbative input, should 
eventually predict the effective ${k}_{T}$ values expected for each 
process and \s. We will employ \avkt\ values representative of the 
kinematic distributions in the data, and based on comparisons with the 
same model as used to modify the NLO inclusive cross sections.

	The effects of soft-gluon radiation are also included in 
QCD Monte Carlo programs such as {\sc Pythia} \cite{pythia}, 
{\sc Isajet} \cite{isajet}, and {\sc Herwig} \cite{herwig}.
However, in these programs the emission is normally cut off at a 
relatively high parton virtuality,  with the remaining ${k}_{T}$ effect 
supplied by a Gaussian smearing similar to that discussed above. 
For fixed-target energies, essentially all of the ${k}_{T}$ effects are 
provided by this phenomenological Gaussian term. The above programs  
differ in the details of the way parton energy and momentum are rescaled after 
${k}_{T}$ is inserted, which can also produce quantitative differences
in results.   

\section*{Applications of the \kt\ Model to Data}

The experimental consequences of \kt\ smearing are expected to
depend on the collision energy. At the Tevatron collider, the smallest photon
$p_T$ values probed by the CDF and \DZERO\ experiments are rather large
(10--15 GeV/$c$), and the \kt-enhancement factors modify only the 
very lowest end of the \pt\ spectrum, where $p_T$ is not significantly 
larger than \kt. In the energy range of the E706 measurements, large \kt-effects 
can modify both the normalizations and the shapes of the cross sections as 
functions of \pt. Consequently, E706 data provide a particularly sensitive 
test of the \kt\ model. At lower fixed-target energies, the \kt\ enhancements are
expected to have less \pt\ dependence over the range of available measurements,
and can therefore be masked more easily by uncertainties in experimental 
normalizations and/or choices of theoretical scales. Nonetheless, the UA6 and WA70 
data generally support the expectations from \kt\ smearing.

\subsection*{Comparisons to Tevatron Collider Data}

\begin{figure}[tbp]
\begin{center}
\epsfxsize=14cm
\epsfysize=17.5cm
\mbox{\epsfbox{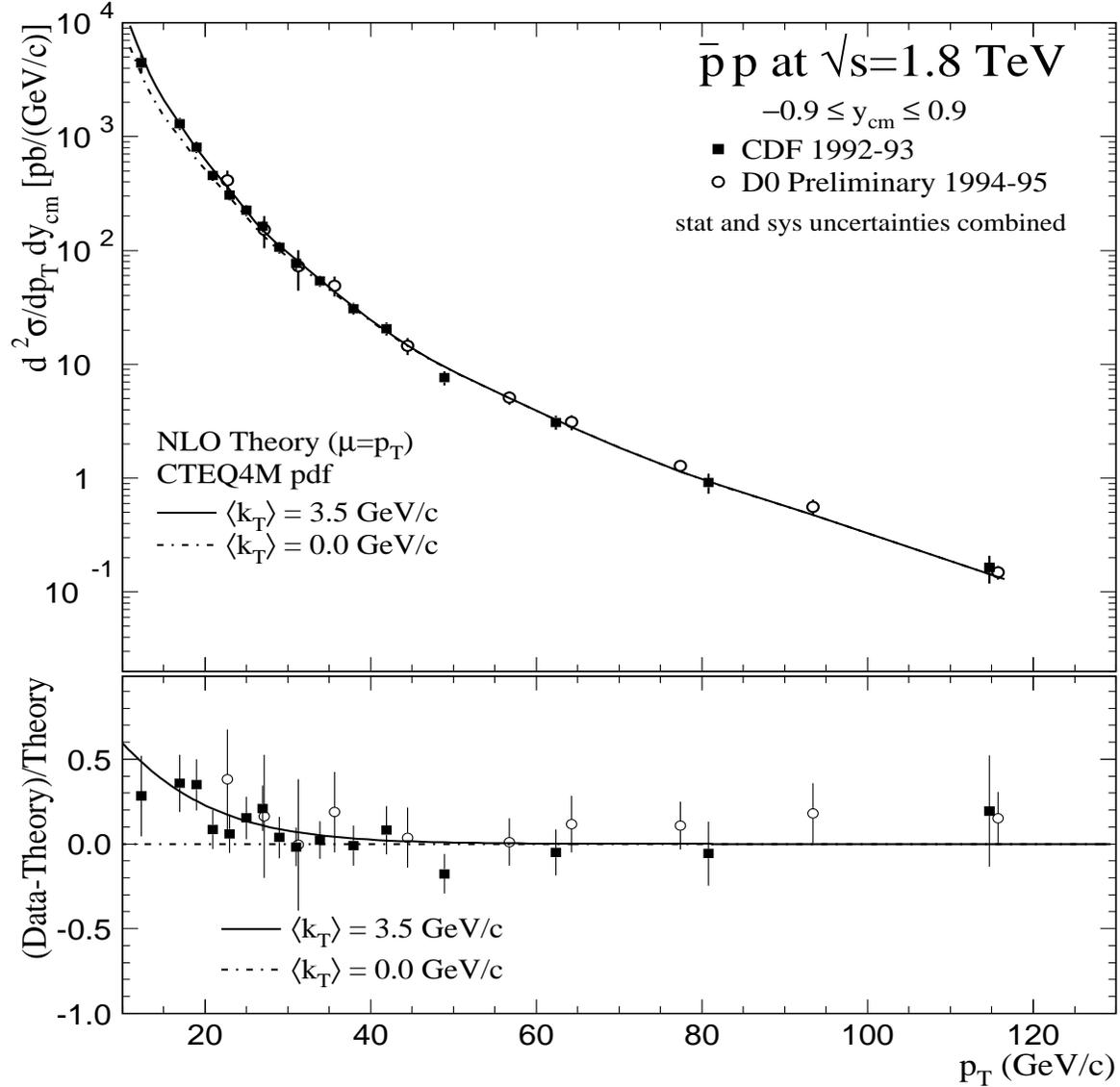}}
\end{center}
\caption{
Top: The CDF and \DZERO\ isolated direct-photon cross sections, compared to
NLO theory without \kt\ (dashed) and with \kt\ enhancement for 
\avkt\ = 3.5 GeV/$c$ (solid), as a function of $p_T$.
Bottom: The quantity (Data--Theory)/Theory 
(for theory without \kt\ adjustment),
overlaid with the expected effect from ${k}_{T}$ enhancement
for \avkt\ = 3.5 GeV/$c$. 
The error bars have experimental statistical and systematic uncertainties
added in quadrature.
} 
\label{fig:cdfd0}
\end{figure}

	At the Tevatron collider, the above model of soft-gluon radiation
leads to a relatively simple modification of the NLO cross section.
In Fig.~\ref{fig:cdfd0} we compare the CDF and \DZERO\ isolated direct-photon 
cross sections \cite{cdfd0phot} to theoretical NLO calculations with and
without \kt\ enhancement.
In the lower part of the plot we display the quantity (Data--Theory)/Theory;
for the collider regime we did not multiply the NLO theory
by the \kt-enhancement factor, but instead displayed the full deviation
of the NLO calculation from the data. The expected effect from  ${k}_{T}$ 
enhancement is also shown for \avkt\ = 3.5 GeV/$c$. 
This is the approximate value of \avkt\ per parton  measured in diphoton 
production at the Tevatron \cite{cdfd0phot}, and one expects a similar
\avkt\ per parton for single-photon production. 
(In the diphoton process, the 4-vectors of the photons can be 
measured precisely, providing a direct determination of the transverse
momentum of the diphoton system, and thereby \avkt.)

As seen in Fig.~\ref{fig:cdfd0}, the \kt\ effect diminishes rapidly 
with \pt\ and is essentially negligible above  $\approx$30 GeV/$c$. 
The trend of deviations of NLO calculations from the measured inclusive 
cross sections is described reasonably well by the expected ${k}_{T}$ effect. 
Some of the observed excess can be attributed to the fragmentation contribution
to isolated direct-photon production \cite{gordon}, but this alone
cannot account for the entire deviation of the theory from data.

The larger statistics in the Tevatron collider Run IB samples (currently under 
analysis) will allow for a more detailed examination of the low-${p}_{T}$ behavior 
of the photon cross section. (The CDF data included in the plot are from 
Run IA only.) A similar enhancement is expected for jet production at low \pt,
but larger  experimental uncertainties,
and the relatively large additional non-perturbative effects 
expected in this region \cite{xt}, preclude a useful comparison.

\subsection*{Comparisons to E706 Data}

\begin{figure}[tbp]
\begin{center}
\epsfxsize=14cm
\epsfysize=17.5cm
\mbox{\epsfbox{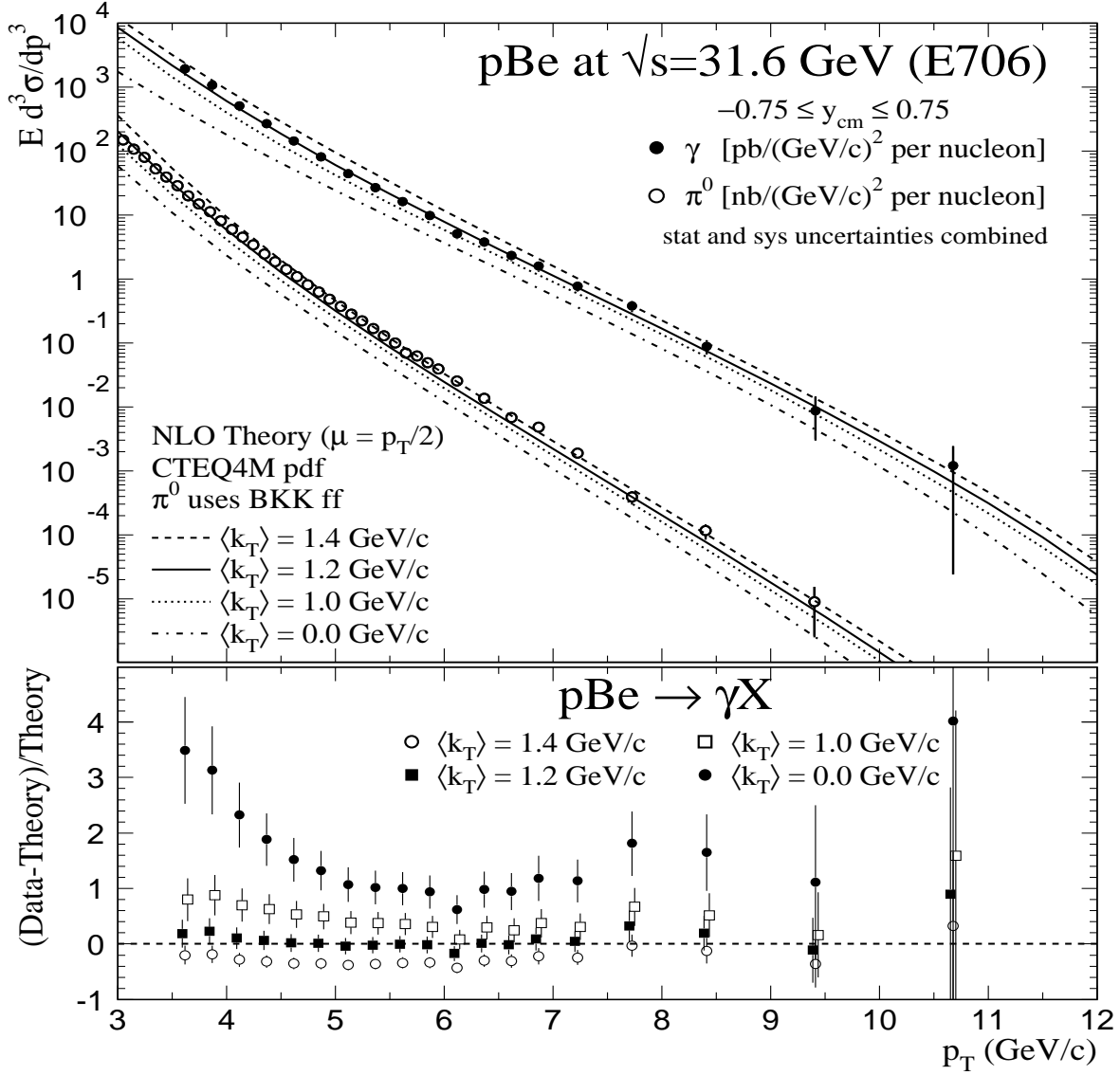}}
\end{center}
\caption{
Top: The photon and $\pi^0$ cross sections from E706 at \s\ = 31.6 GeV compared to
${k}_{T}$-enhanced NLO calculations.
Bottom: The quantity (Data--Theory)/Theory for
direct-photon production, using ${k}_{T}$-enhanced NLO calculations for several
values of \avkt.
The error bars have experimental statistical and systematic uncertainties
added in quadrature.
The points corresponding to calculations with different \avkt\ are slightly
staggered in \pt, to reduce the overlap of experimental error bars.
} 
\label{fig:xs530}
\end{figure}

\begin{figure}[tbp]
\begin{center}
\epsfxsize=14cm
\epsfysize=17.5cm
\mbox{\epsfbox{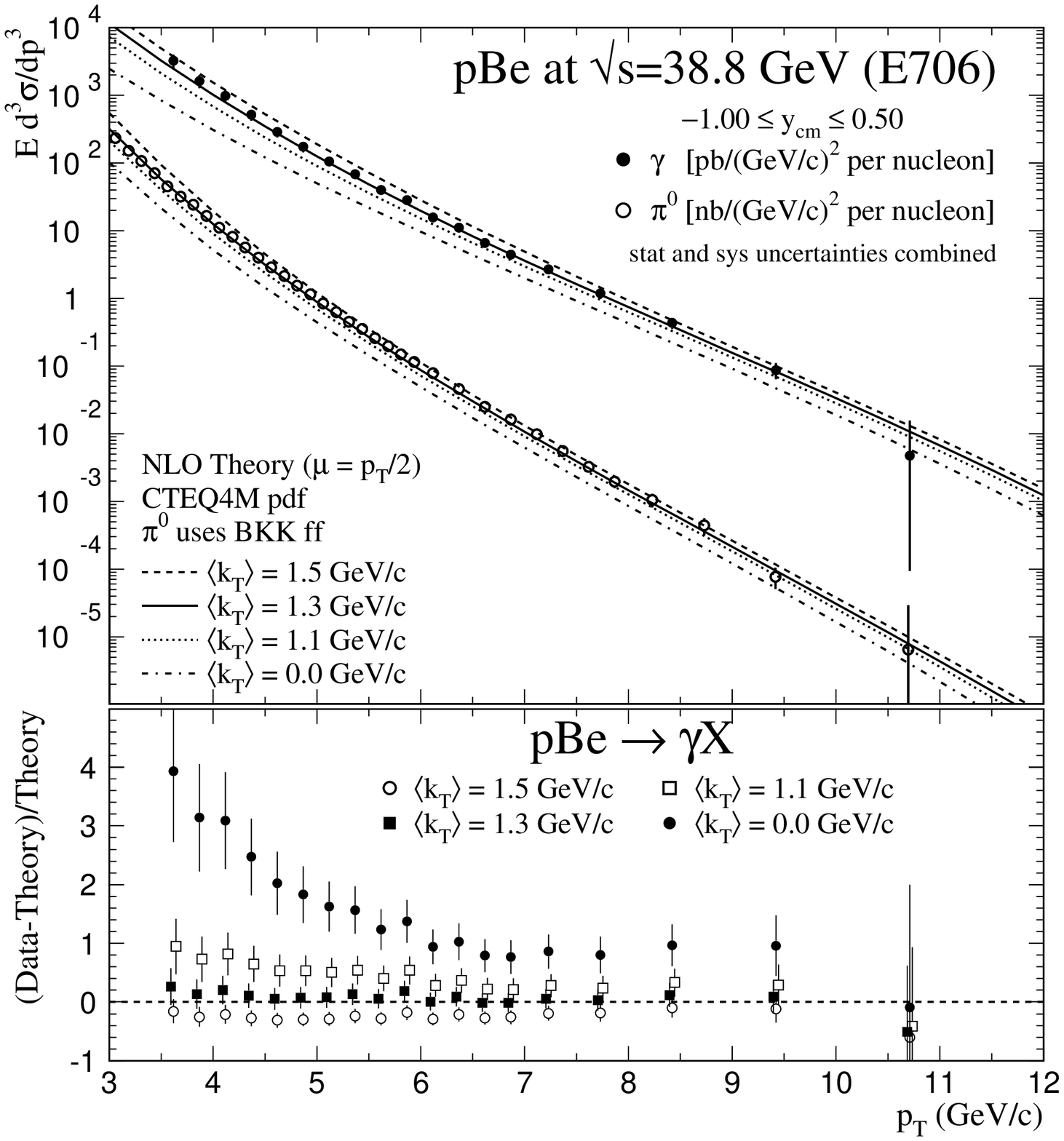}}
\end{center}
\caption{
Top: The photon and $\pi^0$ cross sections from E706 at \s\ = 38.8 GeV compared to
${k}_{T}$-enhanced NLO calculations.
Bottom: The quantity (Data--Theory)/Theory for
direct-photon production, using ${k}_{T}$-enhanced NLO calculations for several
values of \avkt.
The error bars have experimental statistical and systematic uncertainties
added in quadrature.
} 
\label{fig:xs800}
\end{figure}

\begin{figure}[tbp]
\begin{center}
\epsfxsize=14cm
\epsfysize=17.5cm
\mbox{\epsfbox{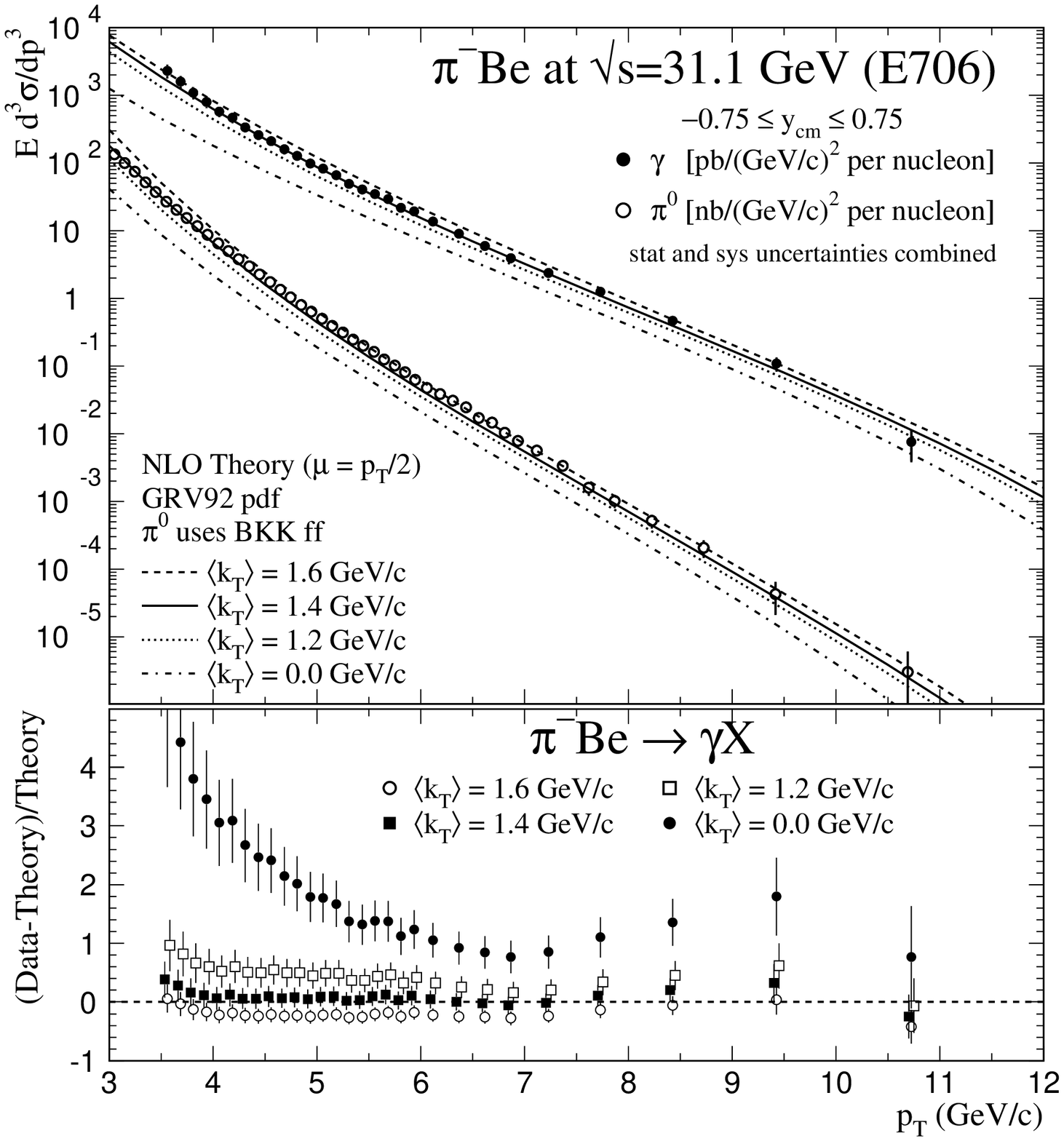}}
\end{center}
\caption{
Top: The photon and $\pi^0$ cross sections from E706 at \s\ = 31.1 GeV 
for incident $\pi^-$ beam, compared to ${k}_{T}$-enhanced NLO calculations.
Bottom: The quantity (Data--Theory)/Theory for direct-photon production, 
using ${k}_{T}$-enhanced NLO calculations for several values of \avkt. 
The error bars have experimental statistical and systematic uncertainties
added in quadrature.
} 
\label{fig:xs515}
\end{figure}

The conventional NLO calculations yield cross sections that are signficantly 
below the E706 direct-photon and $\pi^0$ measurements \cite{e706} 
(see Figs.~\ref{fig:xs530},~\ref{fig:xs800}, and~\ref{fig:xs515}). 
No choices of current parton distributions, or conventional
PQCD scales provide an adequate description of the data
(for the comparisons presented here all QCD scales have been set to \pt/2). 
The previously described ${k}_{T}$-enhancement algorithm was used to 
incorporate the effects of soft-gluon radiation in the 
calculated yields. That is, the theory results plotted in the figures 
represent the NLO calculations multiplied by ${k}_{T}$-enhancement factors 
${K}$(${p}_{T}$) \cite{Be}.

Because parton distributions for nucleons are known better than those for pions,
we first present comparisons of the various model calculations with proton-beam 
data. As seen at the bottoms of Figs.~\ref{fig:xs530} and~\ref{fig:xs800}, 
the NLO theory, when supplemented with appropriate ${k}_{T}$
enhancements, is successful in describing both the shape and normalization
of the E706 direct-photon cross sections at both \s\ = 31.6 GeV and 38.8 GeV.
As expected, the ${k}_{T}$-enhancement factors affect the normalization of 
the cross sections, as well as the shapes of the \pt\ distributions. 
The values of \avkt\ = 1.2 GeV/$c$ at $\sqrt{s}$ = 31.6 GeV, and 1.3 GeV/$c$ 
at $\sqrt{s}$ = 38.8 GeV, provide good representations of the incident-proton 
data. Both ${k}_{T}$ values are consistent with those emerging  from a comparison 
of the same PQCD Monte Carlo with E706 data on the production of high-mass 
$\pi^0\pi^0$, $\gamma\pi^0$ and $\gamma\gamma$ pairs \cite{e706}.

In Fig.~\ref{fig:xs515} we present comparisons between calculations and 
E706 data for $\pi^-$Be interactions at \s\ = 31.1 GeV. Here again, 
the uncorrected (\avkt\ = 0) NLO theory is not consistent with the data.
Once the \kt-enhancement factors are applied, good agreement is observed 
between data and calculations for \avkt$\approx$1.4 GeV/$c$. (Note that 
DY data lead one to expect a higher value of \avkt\ for a $\pi^-$
beam than for a proton beam of the same energy.)

\begin{figure}[tb]
\begin{center}
\epsfxsize=9cm
\epsfysize=12cm
\vskip-1.cm
\mbox{\epsfbox{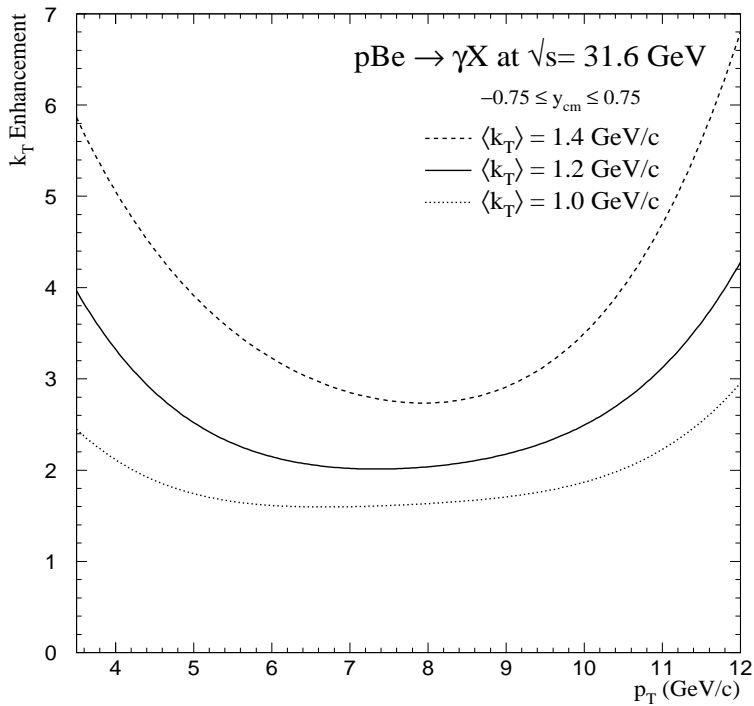}}
\vskip-1.cm
\end{center}
\caption{
The variation of ${k}_{T}$ enhancements, $K(\mpt)$,
relevant to the E706 direct-photon 
data for protons at \s\ = 31.6 GeV, for different values of \avkt.
} 
\label{fig:kfac530}
\end{figure}

For comparison, results of calculations using \avkt\ values $\pm$0.2 GeV/$c$
of the central values are also shown in the figures.  These
can be taken as an indication of uncertainties on \avkt\ (see next section).
The corresponding enhancement factors $K(\mpt)$ at $\sqrt{s}$ = 31.6 GeV 
are displayed in Fig.~\ref{fig:kfac530}.

	It is interesting to note that, in this energy range, the ${k}_{T}$ 
enhancement does not decrease with increasing ${p}_{T}$ (as for the case 
of the collider), and, in fact, increases at the  highest values of ${p}_{T}$.  
This is a consequence of the wide \x-range spanned by E706, and 
can be understood through the following argument.
At low ${p}_{T}$,  a \avkt\  of 1.2 GeV/$c$ is non-negligible in 
comparison to the \pt\ in the hard-scattering, and the addition of
${k}_{T}$ smearing therefore increases the size of the cross section 
(and steepens the slope). At high ${p}_{T}$ (corresponding to large \x), 
the unmodified NLO cross section becomes 
increasingly steep (due to the rapid fall in parton densities),
and hence the effect of ${k}_{T}$ smearing again becomes larger.

	NLO calculations for $\pi^0$ production have a greater 
theoretical uncertainty than those for direct-photon cross sections 
since they involve parton fragmentation. However, the ${k}_{T}$ effects 
in $\pi^0$ production can be expected to be generally similar to those 
observed in direct-photon production, and the $\pi^0$ data can be used 
to extend tests of the consequences of \kt\ smearing.
Figures~\ref{fig:xs530},~\ref{fig:xs800}, and~\ref{fig:xs515} 
also show comparisons between NLO calculations \cite{aversa} and $\pi^0$ 
production from E706, using BKK fragmentation functions (ff) \cite{BKK}.
The previously described Monte Carlo program was employed to generate 
${k}_{T}$-enhancement factors for $\pi^0$ cross sections, and \avkt\ per 
parton values similar to those that resulted in good agreement 
with direct-photon data also provide a reasonable description of $\pi^0$ 
production. For $\pi^0$ production, an additional smearing 
of the transverse momentum, expected from jet fragmentation,
has been taken into account.

\subsection*{Comparisons to WA70 and UA6 Data}

\begin{figure}[tb]
\begin{center}
\epsfxsize=9cm
\epsfysize=12cm
\vskip-1.cm
\mbox{\epsfbox{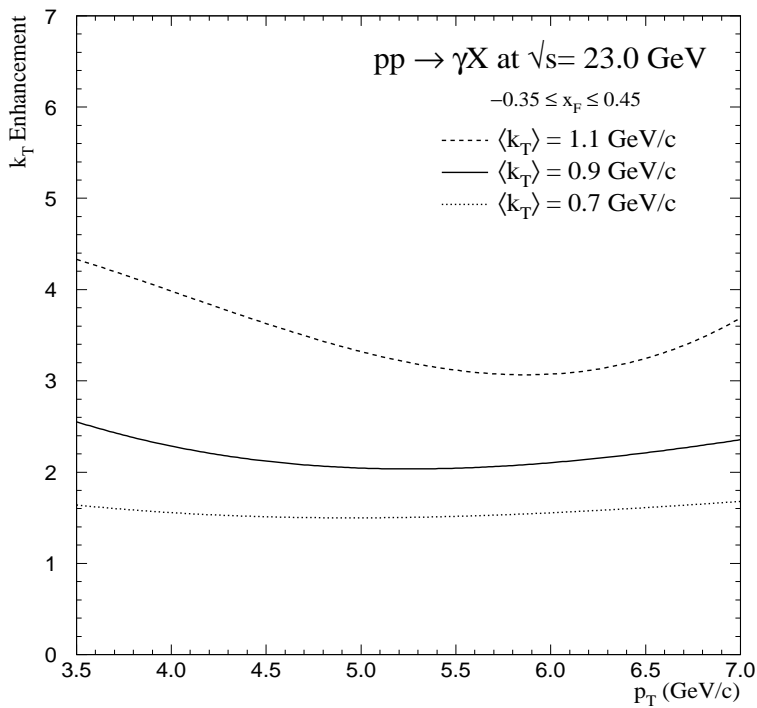}}
\vskip-1.cm
\end{center}
\caption{
The ${k}_{T}$-enhancement factors for direct-photon production 
expected for WA70 $pp$ data.
} 
\label{fig:kfac280}
\end{figure}

We have examined the expectations for the size 
of soft-gluon effects for fixed-target experiments WA70 and UA6.
Both experiments have measured direct-photon production with 
good statistics, and their data have been
included in recent global fits to parton distributions.
WA70 measured direct-photon and $\pi^0$ production in 
$pp$ and $\pi^-p$ collisions at $\sqrt{s}$ = 23.0 GeV \cite{wa70phot}, and
UA6 has recently released \cite{ua6} their final results (with 
substantially reduced uncertainties) for direct-photon production
in $pp$ and $\overline{p}p$ collisions at \s\ = 24.3 GeV.
These center of mass energies are somewhat smaller than 
those of E706, and the \avkt\ values are therefore expected to be smaller
(perhaps of the order of 0.7--0.9 GeV/$c$, based on Fig.~\ref{fig:pairpt}). WA70 has 
compared kinematic distributions observed in diphoton events (for $\pi^- p$ 
interactions) to NLO predictions, and has found that smearing the NLO theory 
with an additional \avkt\ of 0.9$\pm$0.2 GeV/$c$ provides agreement 
with their data \cite{wa70diphot}.
We therefore use this \avkt\ as the central value for the
\kt-enhancement factors for both experiments, and vary the \avkt\ by
$\pm0.2$ GeV/$c$, as with E706. The corresponding ${k}_{T}$ 
enhancement expected over the range of measurements is shown in 
Fig.~\ref{fig:kfac280}. Over this narrower ${p}_{T}$ range, the effect 
of ${k}_{T}$ is essentially to produce a shift in normalization.

\begin{figure}[tbp]
\begin{center}
\epsfxsize=14cm
\epsfysize=17.5cm
\mbox{\epsfbox{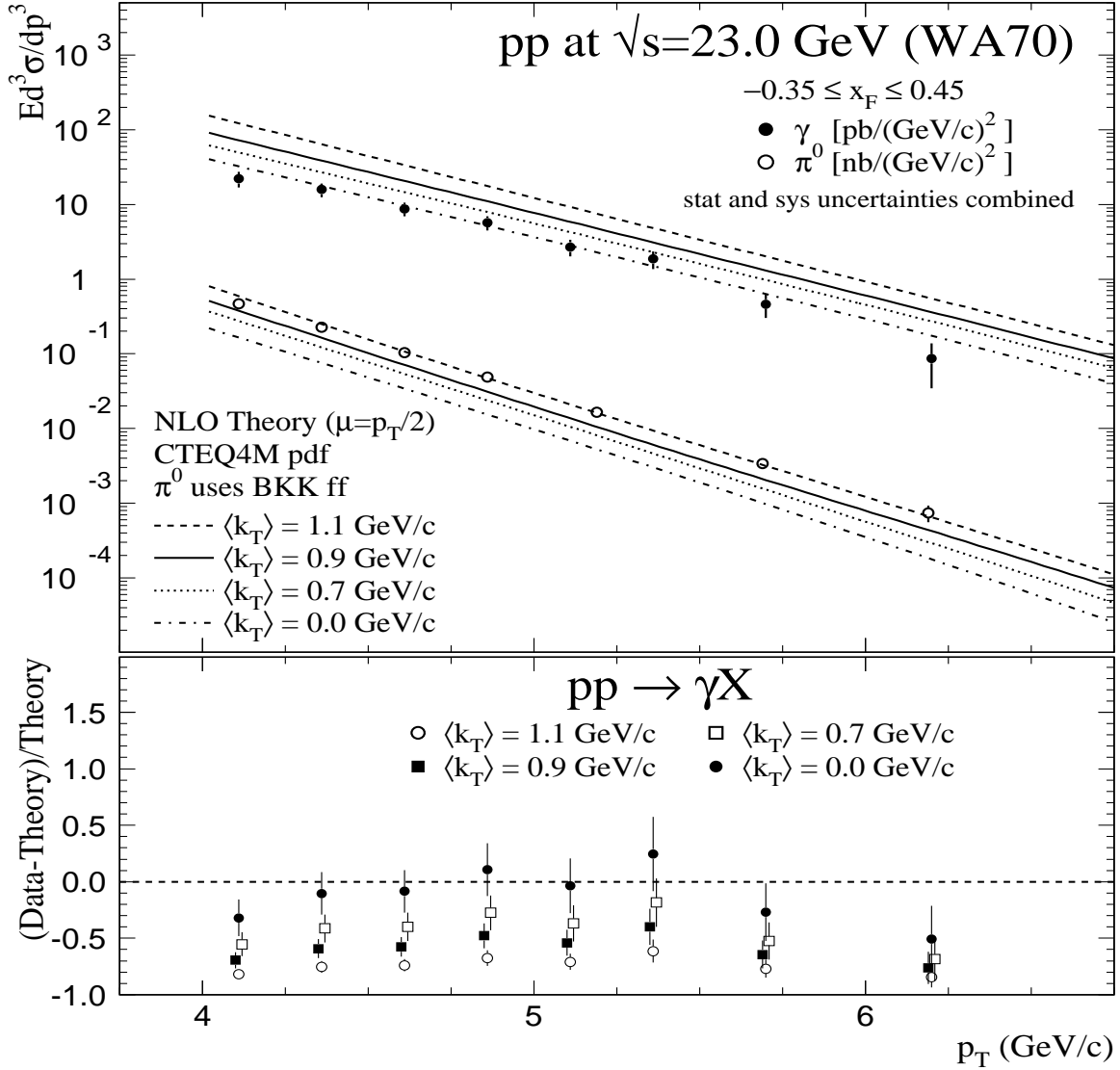}}
\end{center}
\caption{
Top: The photon  and ${\pi}^{0}$ cross sections from WA70 at 
\s\ = 23.0 GeV for incident protons compared to ${k}_{T}$-enhanced 
NLO calculations.
Bottom: The quantity (Data--Theory)/Theory for
direct-photon production, using ${k}_{T}$-enhanced NLO calculations for several
values of \avkt. 
The error bars have experimental statistical and systematic uncertainties
added in quadrature.
} 
\label{fig:xs280p}
\end{figure}

\begin{figure}[tbp]
\begin{center}
\epsfxsize=14cm
\epsfysize=17.5cm
\mbox{\epsfbox{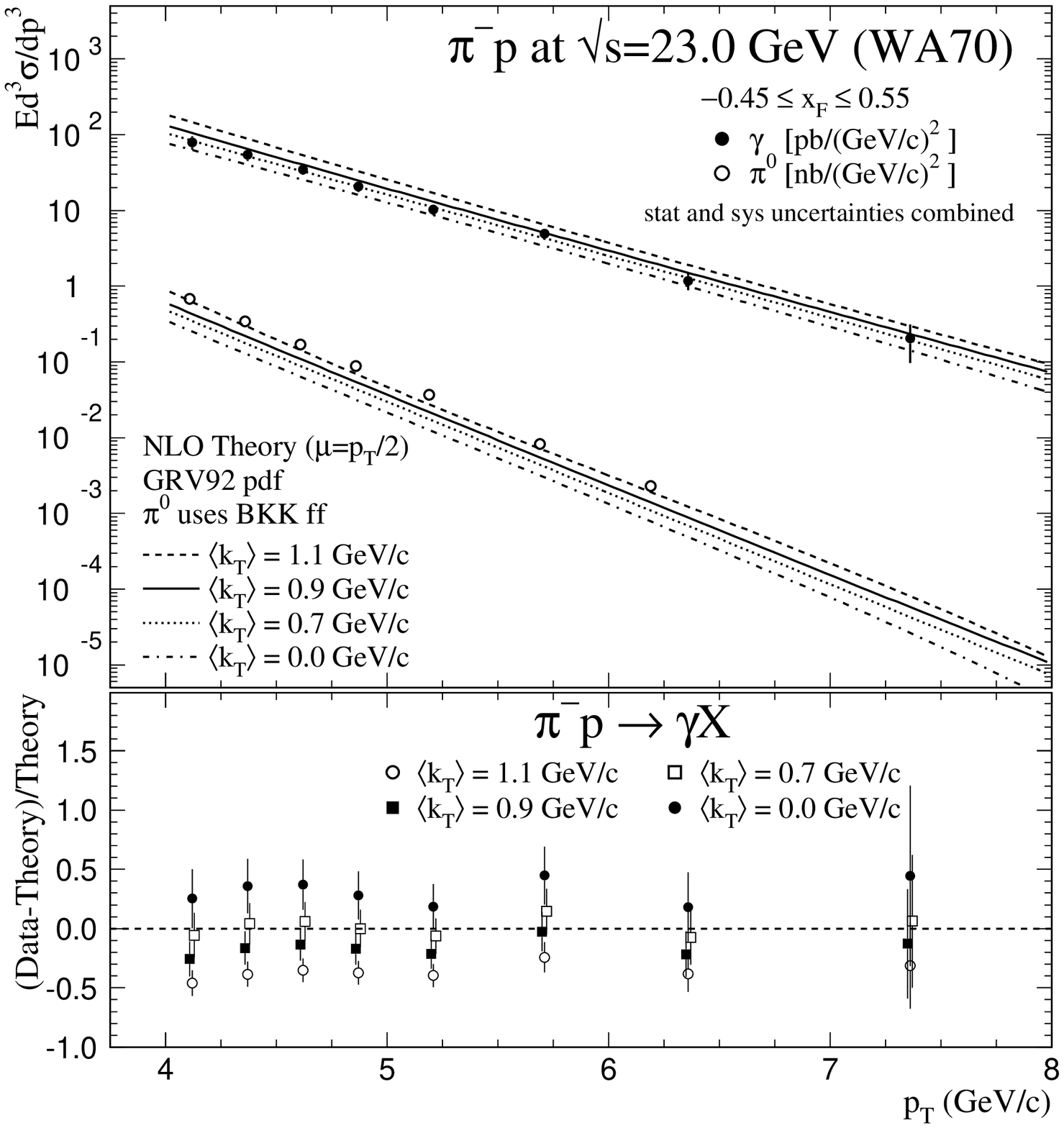}}
\end{center}
\caption{
Top: The photon  and ${\pi}^{0}$ cross sections from WA70 at 
\s\ = 23.0 GeV for
incident ${\pi}^{-}$ beam, compared to
${k}_{T}$-enhanced NLO calculations.
Bottom: The quantity (Data--Theory)/Theory for
direct-photon production, using ${k}_{T}$-enhanced NLO calculations for several
values of \avkt. 
The error bars have experimental statistical and systematic uncertainties
added in quadrature.
} 
\label{fig:xs280pim}
\end{figure}

Comparisons of the WA70 direct-photon and ${\pi}^{0}$ 
cross sections with the ${k}_{T}$-enhanced NLO calculations are
shown in Figs.~\ref{fig:xs280p} and~\ref{fig:xs280pim}.
Renormalization and factorization scales of ${p}_{T}$/2 are 
used in the NLO calculations, as in the E706 comparisons.
The ${\pi}^{0}$ cross sections both for incident proton and ${\pi}^{-}$ 
beams, and the photon data from incident ${\pi}^{-}$ beam,
all lie above the NLO calculations for \avkt = 0, and are in better agreement
with the ${k}_{T}$-enhanced calculations; only the photon cross section for
incident protons seems not to require a \kt\ correction.

\begin{figure}[tbp]
\begin{center}
\epsfxsize=14cm
\epsfysize=17.5cm
\mbox{\epsfbox{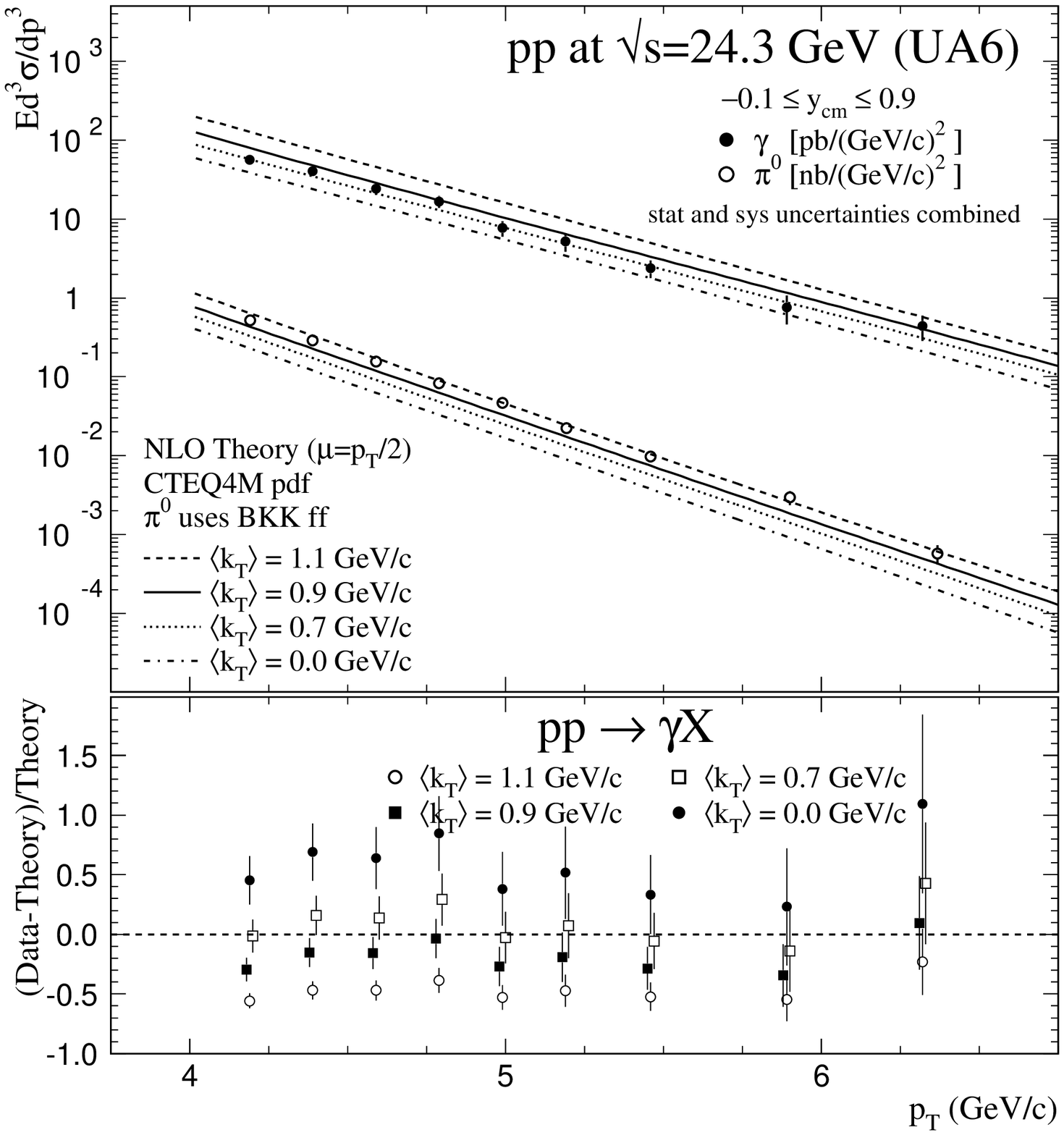}}
\end{center}
\caption{
Top: The photon and $\pi^0$ cross sections from UA6 at \s\ = 24.3 GeV for an
incident proton beam compared to ${k}_{T}$-enhanced NLO calculations.
Bottom: The quantity (Data--Theory)/Theory for
direct-photon production, using ${k}_{T}$-enhanced NLO calculations for several
values of \avkt.
The error bars have experimental statistical and systematic uncertainties
added in quadrature.
} 
\label{fig:xs315p}
\end{figure}

\begin{figure}[tbp]
\begin{center}
\epsfxsize=14cm
\epsfysize=17.5cm
\mbox{\epsfbox{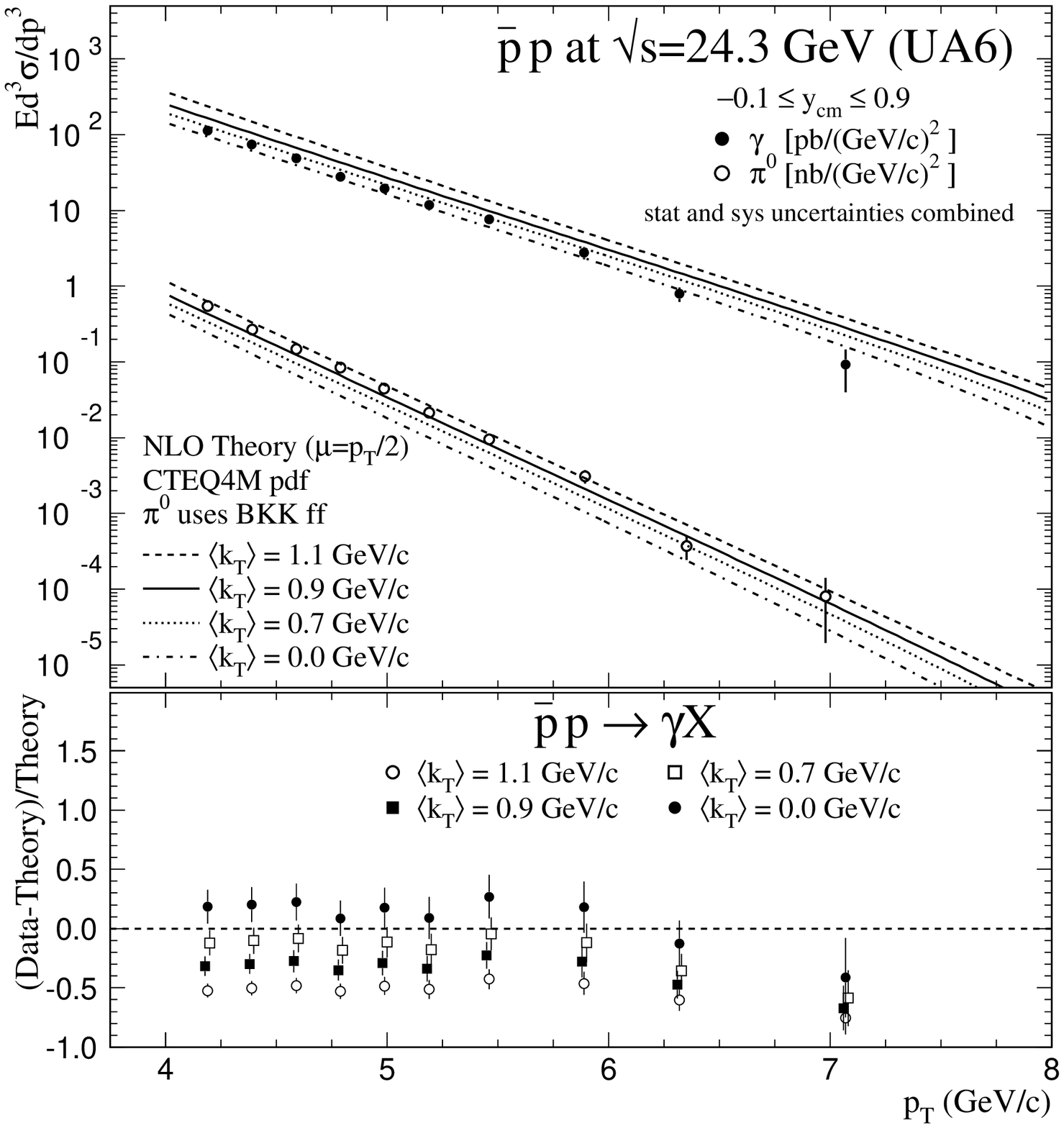}}
\end{center}
\caption{
Top: The photon and $\pi^0$ cross sections from UA6 at \s\ = 24.3 GeV for an
incident antiproton beam compared to ${k}_{T}$-enhanced NLO calculations.
Bottom: The quantity (Data--Theory)/Theory for
direct-photon production, using ${k}_{T}$-enhanced NLO calculations for several
values of \avkt.
The error bars have experimental statistical and systematic uncertainties
added in quadrature.
} 
\label{fig:xs315pbar}
\end{figure}

The latest photon cross sections from UA6  for $pp$ and $\overline{p}p$ 
scattering are shown in Figs.~\ref{fig:xs315p} and~\ref{fig:xs315pbar}.
The photon cross section for $pp$ interactions lies clearly above the 
NLO calculation for \avkt\ = 0, but is consistent with \kt-adjusted
calculations for \avkt\ in the range of 0.7--0.9 GeV/$c$. The result
for $\overline{p}p$ interactions is also above the unmodified NLO calculation,
but requires a smaller value of \avkt. We note that the dominant
production mechanisms for the two processes are different: quark-gluon
Compton scattering dominates for $pp$, and  $\overline{q}q$
annihilation for $\overline{p}p$ at the UA6 energy.
As in the case of E706 and WA70 measurements, the UA6 $\pi^0$ cross sections
are also higher than the NLO calculation without \kt, and can be much better
described by introducing \kt\ enhancement.

\section*{Discussion of \kt-Smearing Procedures}

In this section, we consider uncertainties in the Monte Carlo method 
employed in our previous discussion, and comment on an analytic approach 
for calculating such \kt\ enhancements.
 
\subsection*{Uncertainties in the Monte Carlo Model and Related Issues}

Our approach contains several phenomenological parameters
that affect the range of its results, the most important of which is the 
amount of Gaussian smearing represented by \avkt. Thus far, both
the value of \avkt\ (with its dependence on \s) and its possible range 
of uncertainty can only be determined empirically. A range of variation 
of $\pm$0.2 GeV/$c$ for \avkt$\sim$1--2 GeV/$c$ appears reasonable on the 
basis of several  considerations. These include: (i) the range of \kt\ values 
inferred from the E706 high-mass pair distributions for several kinematic 
variables; (ii) observed differences between photon and $\pi^0$ results;
(iii) comparisons of \avkt\ values required in inclusive cross sections to
those representing the properties of massive pairs at E706 and 
WA70/UA6 energies, and (iv) differences between dimuon, diphoton, 
and dijet values of \avptp. (The dependence of the \kt-enhancement factor
on the variation in \avkt\ was illustrated
in Figs.~\ref{fig:kfac530} and~\ref{fig:kfac280}.)

For the fixed-target experiments discussed in the previous section,
we already presented calculations using \avkt\ values of 0.2 GeV/$c$ above 
and below the selected central values. The observed variation 
in predictions reflects an uncertainty in the 
${k}_{T}$-enhanced theoretical results. The dependence of  
$K(\mpt)$ on \avkt\ is especially strong at the extremes (low 
and high) in ${p}_{T}$. These calculations used  renormalization and 
factorization scales of ${p}_{T}$/2. Changing the scale to ${p}_{T}$
reduces predicted cross sections at fixed-target energies
by about 30--40{\%}, a factor comparable to the 
estimated spread in the ${k}_{T}$ factors over much of the 
${p}_{T}$ range; the full uncertainty in the calculations 
must include contributions from uncertainties in both \avkt\ and 
QCD scales.

	Another parameter in the model, a propagator mass $m$, is
introduced to regularize the divergences of the leading-order QCD matrix
elements due to the propagator factors 
1/$\hat s$, 1/$\hat t$, and 1/$\hat u$ in the configuration where one or
more of these invariants approaches zero.  To avoid the large weights 
associated with these configurations, the propagators are replaced by
1/($\hat s$ + ${m}^2$), 1/($\hat t$ - ${m}^2$), and 1/($\hat u$ - ${m}^2$),
respectively, where $m$ has a typical value of order 1 GeV.  
The physical effect of the propagator mass is to cut off the region where 
almost all of the transverse momentum of the produced photon is due to the
Gaussian smearing, and very little to the hard scattering.
While the \kt-enhancement factor 
is relatively insensitive to the value of $m$ for the measured range at collider 
energies, it is somewhat sensitive for data at fixed-target energies. 

\begin{figure}[tb]
\begin{center}
\epsfxsize=9cm
\epsfysize=12cm
\vskip-1.cm
\mbox{\epsfbox{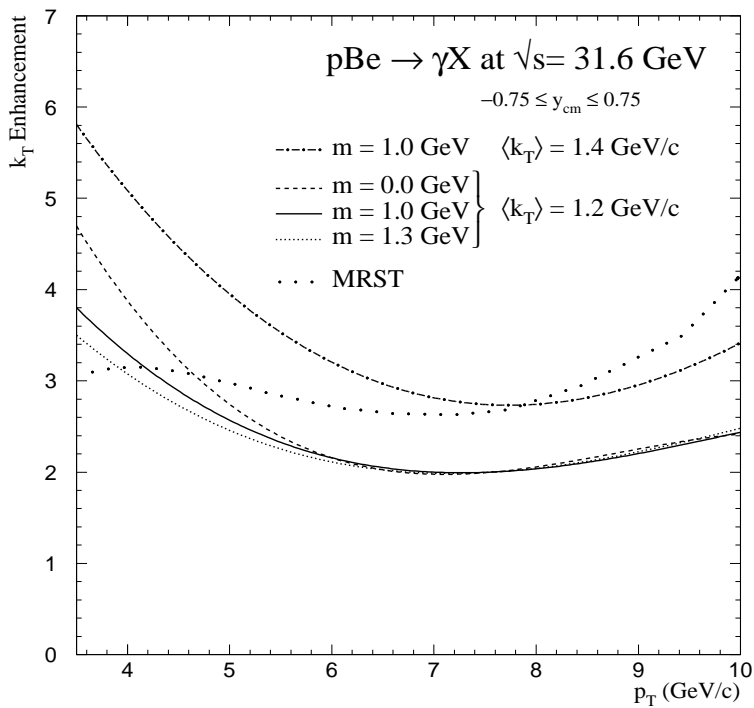}}
\vskip-1.cm
\end{center}
\caption{
The effect of different propagator masses $m$ for predictions of ${k}_{T}$ 
enhancements $K(\mpt)$ for E706 direct-photon data at \s\ = 31.6 GeV. 
The $k_T$ enhancement used by the MRST group {\protect \cite{MRST}}
is also shown in the figure.
} 
\label{fig:propagator}
\end{figure}

To illustrate the sensitivity of the ${k}_{T}$-enhancement factor $K(\mpt)$
to the choice of the propagator mass, we display in Fig.~\ref{fig:propagator} 
the results for $m$ = 1.0 GeV (default value), 1.3 GeV, and 0,
for $p{\rm Be} {\rightarrow} {\gamma}X$ at $\sqrt{s}$ = 31.6 GeV.
Above \pt\ values of 5.5--6.0 GeV/$c$, the \kt\ enhancement has little 
sensitivity to $m$. At lower $p_T$, the dependence on the propagator mass 
should be considered as a measure of an uncertainty of the model in this 
region of phase space. Clearly, larger enhancements are obtained at low 
${p}_{T}$ when there is no propagator mass ($m$ = 0).

The value of \avkt, appropriate in the calculation, depends on the kinematic 
situation. The increase of the value of \avkt\ with $s$ is
understood, in general terms, as the result of an increase in the available phase
space for multiple-soft-gluon emission. This growth is predicted by a CSS-type 
of resummation for Drell-Yan processes \cite{stirling} and, as illustrated in 
Fig.~\ref{fig:pairpt}, has been observed in Drell-Yan and diphoton data.
For simplicity, the model calculation
assumes a constant value of \avkt\ for a given $\sqrt{s}$. In
reality, various physical effects can modify this simple choice and cause
a modification in the shape of the enhancement factor $K(\mpt)$. 
Below, we discuss a few of these effects.

First, one might expect a dependence of \avkt\ on $\hat s$, the hardness of the
partonic interaction, similar to that on $s$. 
A study of diphoton production at $\sqrt{s}$ = 31.5 GeV, using the {\sc Resbos} 
program \cite{resbos}, indicates that the \avkt\ increases from
about 1.2 GeV/$c$ for photons with $p_T$  of 5 GeV/$c$ to approximately 1.4
GeV/$c$ for photons with $p_T$ of 10 GeV/$c$. If direct-photon production were
to have a similar behavior, the anticipated \kt\ enhancement at  a $p_T$ of 
10 GeV/$c$ would be about 40\% higher than for a constant \avkt\ value of 
1.2 GeV/$c$ (see the dependence of the enhancement factor on \avkt\ in
Fig.~\ref{fig:kfac530}). 

Suppression mechanisms may also exist at large \pt\ due to the restriction of phase 
space for gluon emission from large-\x\ partons. A gluon emitted from an incoming 
parton carries away a momentum fraction $\Delta x = 2 p_Te^y/\sqrt{s}$, where $y$ 
is the rapidity of the gluon. Since gluons emitted at forward rapidities would carry 
away more of the incident parton's momentum than available, the effective
rapidity range for gluon emission becomes restricted to central rapidities.
Such effects are not taken into account correctly in the simple  $k_T$ models
discussed above.

Finally, there has been much recent interest in studying the effects of
resumming large logarithms of the form $\ln(1-x_T)$, where 
$x_T = 2\mpt/\sqrt{s}$ \cite{cmn,los}. As $x_T$ approaches 1, for any 
hard-scattering process, the perturbative cross section is enhanced by 
powers of $\ln(1-x_T)$ that have to be resummed at all orders. These 
types of effects should currently be negligible for direct-photon production 
at the Tevatron collider (because data are available only for relatively small 
values of $x_T$), but may be important at fixed-target energies.
Definitive answers to the question of whether this effect leads to a net
enhancement or suppression, and its consequences for the shape of the \pt\
distribution, are not presently available.
A treatment that includes both $k_T$ and threshold resummation effects may be
necessary for a more satisfactory description of the fixed-target data. 

\subsection*{Analytic Smearing Methods}

An alternative to the Monte Carlo calculation of the \kt\ 
enhancement is its implementation through analytic convolution
of the theoretical cross section with a (Gaussian) $k_T$ distribution 
in either one or two dimensions. The latter is a better approximation,
but the correction to the one-dimensional convolution is 
second order in ${k}_{T}/{p}_{T}$, leading to a difference of about 10\% 
for the fixed-target applications. To compare to the Monte Carlo results, 
the parameters of such analytic convolutions
must be interpreted in terms of the parton-level \avkt\ values.

We consider the kinematics 
of the production of a direct photon accompanied by a recoil jet.
As mentioned before, the total transverse momentum imparted to the final
state ($\gamma$+jet) by the colliding partons has an RMS width 
\begin{equation}
\sigma_{2partons,2D} = \sqrt{2} \sigma_{1parton, 2D}.
\end{equation}
However, the single photon is subject only to half of this 
transverse kick, because it
shares the total \pt\ with the recoil jet. 
Consequently, the  ${k}_{T}$ RMS width relevant
for an analytic smearing of the single-inclusive photon cross section is
\begin{equation}
\sigma_{\gamma,2D} = {1 \over 2} \sigma_{2parton, 2D} = {1 \over \sqrt{2}} 
\sigma_{1parton,2D}. \label{eq:inclsmear}
\end{equation}
The above can be illustrated by  a simple example. Consider the production of 
a photon and a jet with equal and opposite ${p}_{T}$ values (for example, 
4 GeV/$c$) close to  90 degrees in the center of mass. Compare this to the 
situation when the colliding partons impart some amount of ${k}_{T}$,
say 1 GeV/$c$, in the direction of the photon. The photon $p_T$ is now 4.5 GeV/$c$, 
and the jet $p_T$ becomes 3.5 GeV/$c$ (in the opposite direction),
resulting from a total of 1 GeV/$c$ of ${k}_{T}$ imparted to the photon+jet system.
Thus, the photon itself only receives half of the partonic total. This factor
of two is critical, and can be easily overlooked, in calculating $k_T$ smearing
using analytic methods. The above conclusions have been verified within the
explicit Monte Carlo treatment of event kinematics. (The convolution formulae 
for $k_T$ smearing are discussed more fully in the Appendix.)

Recently, the MRST group has produced a new set of parton distributions, 
incorporating ${k}_{T}$ effects in their analysis of the WA70 and E706 
direct-photon data \cite{MRST}; they used an analytic smearing technique
rather than a Monte Carlo approach. The correction algorithm differs in detail 
with the one discussed above, and seems to require a significantly 
smaller \avkt\ to describe the data (about 0.65 GeV/$c$ per incoming 
parton as compared to 1.2 GeV/$c$ that was used for E706 at $\sqrt{s}$ = 31.6 GeV).
This discrepancy may be due to a difference in the interpretation of \kt\
that is related to the factor of two present in Eq.~\ref{eq:inclsmear}.
If we reinterpret the \avkt$_{\rm MRST}$ of 0.65 GeV/$c$ as our 1.3 GeV/$c$ per 
parton, then agreement is restored. Nevertheless, the ${k}_{T}$-enhancement 
factors differ somewhat in the two techniques, leading to 
different conclusions about the gluon distribution, especially at large
$x$ (see next section). For comparison, the MRST ${k}_{T}$ factor
for E706 (at $\sqrt{s}$ = 31.6 GeV) has also been plotted in
Fig.~\ref{fig:propagator}; it is similar in size to our calculations 
for \avkt\ values in the range of 1.2--1.4 GeV/$c$, but has a different 
shape in \pt. Until these differences are understood, we must assume that
there is a significant shape-dependent uncertainty associated with the
particular assumptions used for modeling \kt\ effects.

\section*{Impact on the Gluon Distribution}

It is now generally accepted that the uncertainty in the gluon distribution 
at large \x\ is still quite large. Thus, it would appear important to incorporate
further constraints on the gluon, especially from direct-photon data.
In the following, we describe a global pdf fit that employs our \kt\ model 
in the analysis of E706 direct-photon measurements. Since jet production 
at the Tevatron collider is another available constraint for the gluon content 
at large \x, we discuss the consistency between the E706 direct-photon results 
and the CTEQ4HJ gluon distribution (derived using high-$p_T$ jet
data from CDF) \cite{cteq4hj}. 

\subsection*{Application of \kt\ Enhancements in a pdf Fit}

\begin{figure}[tb]
\begin{center}
\epsfxsize=9cm
\epsfysize=12cm
\vskip-1.cm
\mbox{\epsfbox{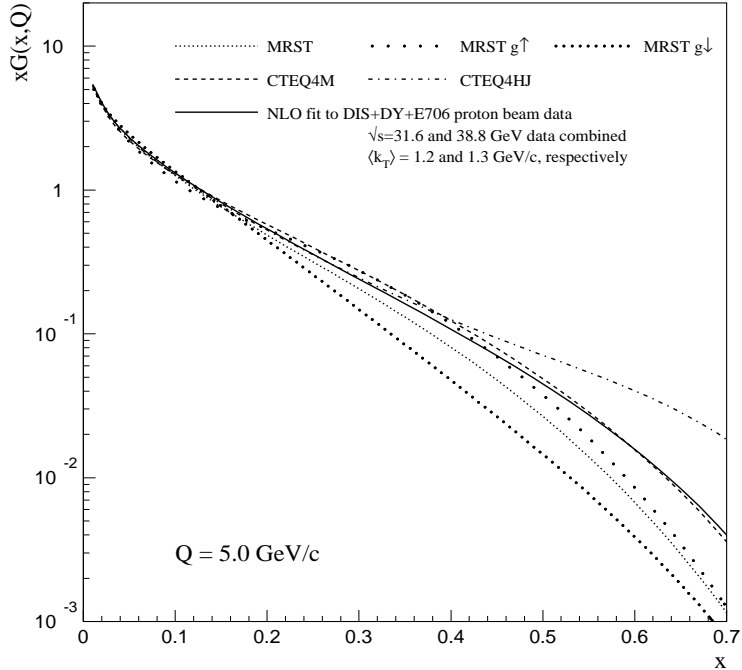}}
\vskip-1.cm
\end{center}
\caption{
A comparison of the  CTEQ4M, MRST, and CTEQ4HJ gluons, 
and the gluon distribution derived from fits that use E706 data.
The $g\!\!\uparrow$ and $g\!\!\downarrow$ gluon densities correspond to
the maximum variation in \avkt\ that MRST allowed in their fits.
} 
\label{fig:gluons}
\end{figure}

To investigate the impact of \kt\ effects on determinations of the gluon 
distribution, we have included the E706 direct-photon cross sections for 
incident protons, along with the DIS and DY data that were used in determining 
the CTEQ4M pdfs, in a global fit to the parton distribution functions.
The CTEQ fitting program was employed to obtain these 
results \cite{cteqfit}, using the NLO PQCD calculations for direct-photon
cross sections, adjusted by the ${k}_{T}$-enhancement factors.
However, the WA70, UA6, CDF, and \DZERO\ data were excluded from
this particular fit.
The resulting gluon distribution, shown in Fig.~\ref{fig:gluons}, is 
similar to CTEQ4M, as might have been expected, since
the ${k}_{T}$-enhanced NLO cross sections using CTEQ4M 
provide a reasonable description of the data
shown  in Figs.~\ref{fig:xs530} and~\ref{fig:xs800}.

The data sets used in determining CTEQ4M did not include the E706 direct-photon 
cross sections, but did use earlier direct-photon data from UA6 and from WA70
(without $k_T$ corrections), along with the inclusive jet cross sections from
CDF and \DZERO\ \cite{cdfd0jet}. The jet cross sections were particularly useful
for defining the gluon distribution in CTEQ4M at moderate values of \x.

 The new MRST gluon distribution (also shown in Fig.~\ref{fig:gluons}) 
is significantly lower than CTEQ4M (and MRSR2) at large \x.
While the MRST fit employs \kt\ enhancements, it attempts to
accommodate the WA70 incident-proton direct-photon data, 
which does not exhibit an obvious \kt\ effect. In addition, the MRST 
\kt-enhancement factors are larger at large \pt\ than our corresponding 
results, resulting in a smaller gluon at large \x.
This further serves to illustrate the extent to which the extracted gluon
distribution is affected by the specific procedures applied in the fit.
In contrast, the CTEQ4HJ specialized gluon distribution (discussed
in more detail in the next subsection) is much larger than CTEQ4M in the
same \x\ range. The current spread of the solutions for the gluon
distribution at large \x\ is uncomfortably large, and additional effort 
is required to resolve these discrepancies.

\subsection*{Discussion of the CTEQ4HJ Gluon}

As presented above, when analyzed with $k_T$-enhancement factors, the E706 
direct-photon data
lead to a gluon distribution similar to that in CTEQ4M. The implications for
the size of the gluon at large \x\ are especially important because of
the excess observed by CDF in the high-$p_T$ inclusive jet cross section
(when compared to calculations using conventional pdfs).
The CTEQ collaboration produced a global fit (CTEQ4HJ) \cite{cteq4hj}
to improve the description of the high-${p}_{T}$ jet data from CDF in Run IA
\cite{cdfjet}. The high-${p}_{T}$ data points were given an enhanced 
weight to emphasize them in that fit. 
The resulting CTEQ4HJ parton distributions produce a jet cross 
section that by design follows the CDF data points more closely than the 
cross section obtained using CTEQ4M.

The CDF inclusive jet cross section from Run IB \cite{cdfjet1b}
demonstrates an excess at high $p_T$ similar to that observed in Run IA. 
For \DZERO, the inclusive jet cross section from Run IA+IB is consistent
with the NLO QCD calculation using conventional parton distributions
such as CTEQ4M, but can also be well-described with calculations using 
the CTEQ4HJ parton distribution functions; in fact, calculations 
using CTEQ4HJ result in better $\chi^2$ agreement with the \DZERO\ jet
data \cite{blazey}. 

\begin{figure}[p]
\begin{center}
\epsfxsize=14cm
\epsfysize=17.5cm
\mbox{\epsfbox{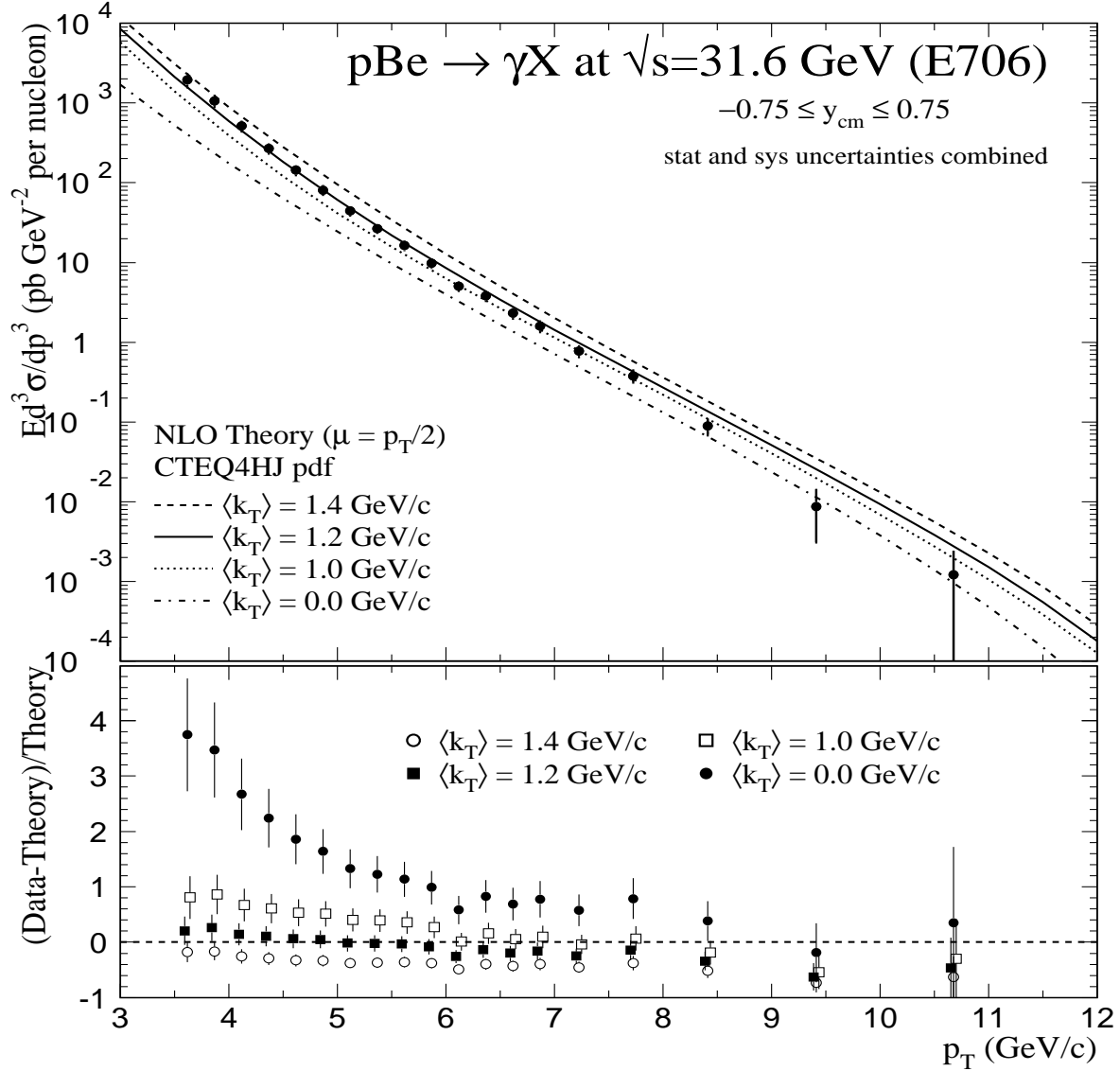}}
\end{center}
\caption{
Top: The photon cross section from E706 at \s\ = 31.6 GeV compared to
${k}_{T}$-enhanced NLO calculations using the CTEQ4HJ parton distribution
functions. 
Bottom: The quantity (Data--Theory)/Theory
using ${k}_{T}$-enhanced NLO calculations for several values of \avkt.
The error bars have experimental statistical and systematic uncertainties
added in quadrature.
} 
\label{fig:xs530hj}
\end{figure}

\begin{figure}[p]
\begin{center}
\epsfxsize=14cm
\epsfysize=17.5cm
\mbox{\epsfbox{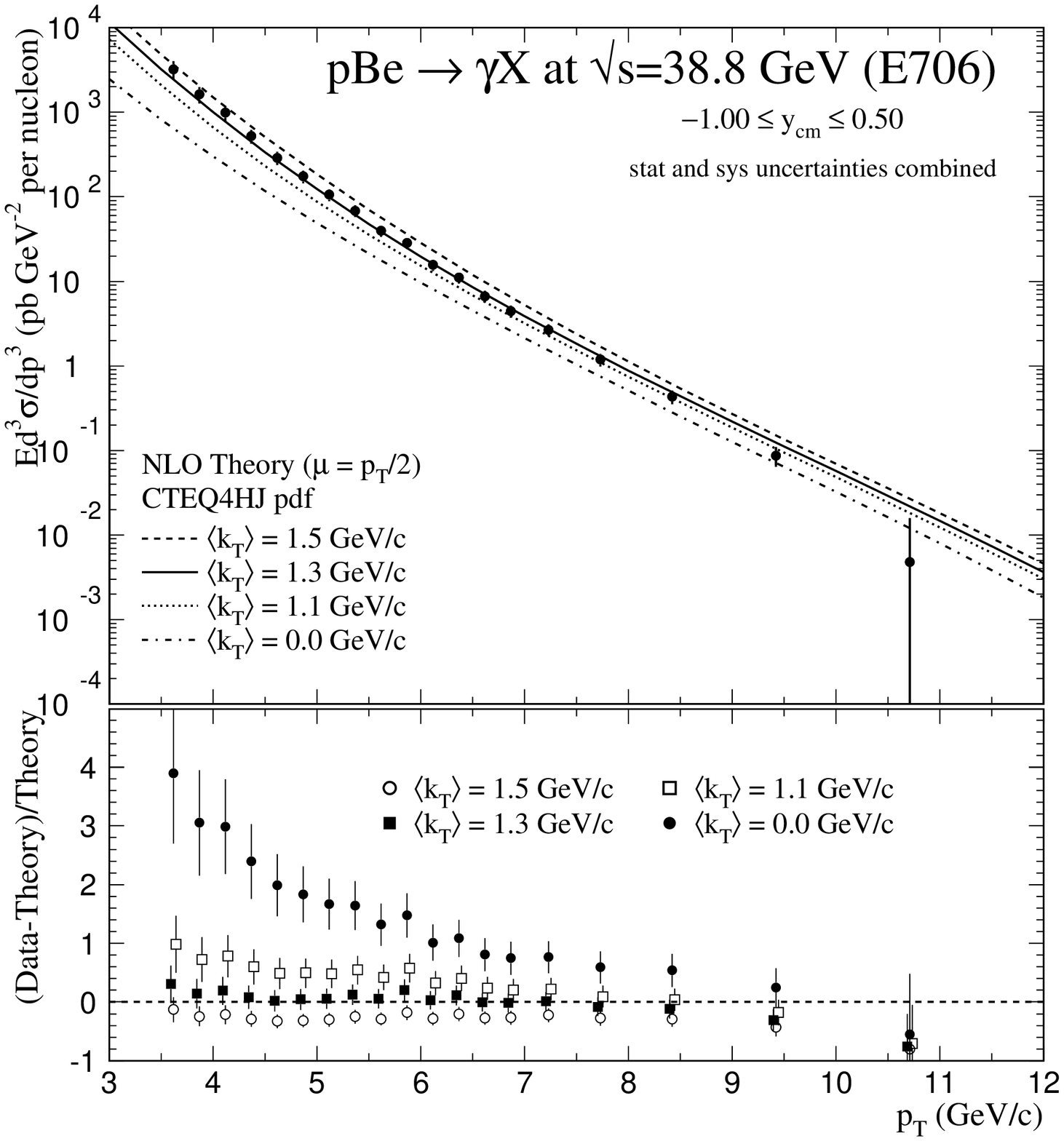}}
\end{center}
\caption{
Top: The photon cross section from E706 at \s\ = 38.8 GeV compared to
${k}_{T}$-enhanced NLO calculations using the CTEQ4HJ parton distribution
functions.
Bottom: The quantity (Data--Theory)/Theory
using ${k}_{T}$-enhanced NLO calculations for several values of \avkt.
The error bars have experimental statistical and systematic uncertainties
added in quadrature.
} 
\label{fig:xs800hj}
\end{figure}

The CTEQ4HJ quark distributions are similar to those obtained in a more 
standard fit (as for example CTEQ4M), since the DIS and DY data provide 
significant constraints on the quark distributions over all \x\
(and on the gluon distribution at small \x) \cite{cteqgluon}.
However, the CTEQ4HJ gluon distribution is 
significantly larger at high \x\ than that of CTEQ4M, by a factor 
of 1.5 at \x\ = 0.5 (for $Q$ = 5 GeV/$c$), and by a factor of 5 at \x\ = 0.7.

Because of the dominance of the $q\overline{q}$ scattering subprocess
in the Tevatron jet cross sections at high ${p}_{T}$, a large change in the 
gluon distribution is required to generate a relatively small change in 
the jet cross section. As the CTEQ exercise has demonstrated, 
until the theoretical issues related to interpretation of direct-photon 
data are resolved, there is freedom within the data sets used in the global 
fits to change the gluon distribution in this way.
It should also be noted that a recent analysis \cite{bodek} of deuteron and proton
structure functions, using corrections for nuclear binding effects in the
deuteron, suggests that the down-quark distribution in the nucleon 
at large \x\ may be significantly larger than previously assumed; clearly this
also influences the calculated cross sections 
for the high-${p}_{T}$ jet production. 

	The CTEQ4HJ gluon distribution is compared to those from CTEQ4M, MRST, 
and the fits including the E706 data in Fig.~\ref{fig:gluons}. 
Figures~\ref{fig:xs530hj} and~\ref{fig:xs800hj} show comparisons of the E706 
direct-photon cross sections and the NLO calculations using the 
CTEQ4HJ parton distributions, with and without the ${k}_{T}$-enhancement factor. 
The shape in \pt\ of these calculations appears less consistent with the data 
than the corresponding results using the CTEQ4M gluon.
However, current uncertainties in the understanding of \kt\ effects
in direct-photon production (discussed in the previous section), 
preclude an unambiguous interpretation of this difference.

\section*{Conclusions} 

We have described a phenomenological model for \kt\ effects in which 
\avkt\ values used in the calculations of \kt-enhancement factors
are derived from data.  Despite uncertainties, 
the results are remarkably successful
in reconciling the data and theoretical calculations for a broad range
of energies. The \kt-enhancement factors improve the agreement
of QCD calculations with E706, UA6, and $\pi^-$ beam WA70 direct-photon 
cross sections over the full \pt\ range of measurements, and at the
low-$p_T$ end of CDF and \DZERO\ results. All fixed-target $\pi^0$
measurements also agree much better with such \kt-enhanced calculations.

The proper treatment of soft-gluon radiation in direct-photon cross
sections can affect the extraction of the gluon distribution,
especially at large values of \x. In this particular treatment of \kt\ 
enhancement, the E706 data, which span the widest \x-range of any 
direct-photon experiment, are in better agreement with the gluon distribution 
from CTEQ4M than from CTEQ4HJ. However, within the  
phenomenological approach, any physical mechanism which gives rise to less  
enhancement at large \x\ than our specific model calculation can make the  
CTEQ4HJ gluon more consistent with the E706 data.

A definitive conclusion regarding the quantitative role of ${k}_{T}$ effects 
in hard scattering, and reliable additional information on the 
large-\x\ gluon distribution, awaits more complete theoretical calculations.
The new generation of direct-photon and inclusive jet measurements 
serves as a strong impetus, and provides a testing ground, for new  
theoretical developments, perhaps incorporating both \kt\ and large-\x\  
resummation, which will have less model dependence than the formulations
discussed in this paper. Further progress built on this interplay between 
theory and experiment will allow a more definitive determination of the 
gluon distribution, especially in the large-\x\ region, where significant 
uncertainties still remain.

\section*{Acknowledgements}

	We thank S. Catani, S. Ellis, E. Kovacs, H.-L. Lai, M. Mangano, 
P. Nason, T. Sj\"ostrand, G. Snow, D. Soper, G. Sterman, J. Stirling, M. Werlen,
and C.-P. Yuan for useful conversations.

\subsection*{Appendix}

In this appendix, we collect formulae relevant to the analytic treatment 
of \kt\ effects in inclusive cross sections. As discussed in the main text, 
we assume a Gaussian description for the parton \kt-distributions;
the width parameters entering the formulae
are labeled explicitly to help keep track of their meaning.

For definiteness, let us consider direct-photon production.
The full 2-dimensional convolution of the (parametrized) differential cross 
section $\Sigma$ (for example, $\Sigma = d\sigma/dp_T$)
with the Gaussian \kt-smearing functions can be written as:
\begin{eqnarray}
\Sigma'(\mpt) = \int d^2k_{T_1} d^2k_{T_2} d^2q_T \;
{1 \over \pi\langle k_{T_1}^2\rangle} e^{-k_{T_1}^2/\langle k_{T_1}^2\rangle} \; 
{1 \over \pi\langle k_{T_2}^2\rangle} e^{-k_{T_2}^2/\langle k_{T_2}^2\rangle} \;
\nonumber \\
\times \Sigma(q_T) \;
\delta^{(2)}(\vec{p}_T-\vec{q}_T-{1 \over 2}(\vec{k}_{T_1}+\vec{k}_{T_2})),
\end{eqnarray}
or, integrating out the $\delta$-function constraints, as
\begin{equation}
\Sigma'(\mpt) = \int d^2{k}_T {1 \over \pi\msgtwo^2} 
          \exp(-\mkt^2/\msgtwo^2) \; \Sigma(| {\vec p}_T - {\vec k}_T |),
\end{equation}
where \sgtwo\ is the width parameter appropriate for the
smearing of the direct-photon inclusive distribution; 
it is related to the width of the parton \kt-distribution  
through Eq.~(\ref{eq:inclsmear}).

The \kt-enhancement factors, defined as $K(\mpt) \equiv \Sigma'(\mpt)/\Sigma(\mpt)$, 
can be calculated numerically. For a more intuitive treatment,
one can simplify the discussion by reducing the above equation
to a 1-dimensional case. Let us decompose the ${\vec k}_T$ vector into 
components parallel and perpendicular to ${\vec p}_T$. Then only the parallel
component strongly affects the value of the photon \pt, while the perpendicular
component affects \pt\ much less for typical \kt$\ll$\pt\ configurations 
(since it adds to \pt\ in quadrature). Neglecting the effect of the 
perpendicular component, and integrating it out, one arrives at a 
1-dimensional approximation:
\begin{equation}
\Sigma'(\mpt) = \int_{-\infty}^{+\infty} dk_T {1 \over \sqrt{2\pi\msg^2}} 
          \exp(-\mkt^2/2\msg^2) \;
          \Sigma(\mpt - \mkt),
\end{equation}
where
\begin{eqnarray}
\msg = {1 \over \sqrt{2}} \msgtwo 
     = {1 \over 2} \msp,
\end{eqnarray}
and we used Eq.~(\ref{eq:inclsmear}) to relate \sgtwo\ and \sp.

Two particularly interesting special cases are the exponential form
of the cross section, used to describe the data at fixed-target
energies, and the $1/\mpt^n$ form, appropriate at colliders.

For an exponential representation of the cross section, $\Sigma \sim
\exp(-b\mpt)$, one obtains 
\begin{equation}
K(\mpt) = \exp({1 \over 8} b^2 \msp^2) = \exp({1 \over 2\pi} b^2 \mavkt^2).
\end{equation}
Using the LO calculation of $pp\rightarrow \gamma X$ at $\sqrt{s}=31.6$ GeV 
(and \avkt\ = 0), one finds $b \approx 1.8 \,  ({\rm GeV}/c)^{-2} $
at \pt\ = 6.5 GeV/$c$ (a value in the middle of the \pt\ range of E706).
The approximate 1-dimensional formula above yields $ K \approx 2.1$ for 
\avkt\ = 1.2 GeV/$c$, compared to $K^{\rm LO} \approx 2.0$ obtained in the full
LO Monte Carlo calculation. In general, despite the approximations in the treatment, 
the analytical results are quite close to the results of the LO calculations. 
The ${1 \over \sqrt{2}}$ factor in the relation of \sgtwo\ and \sp\ 
(Eq. \ref{eq:inclsmear}) is crucial to obtaining this consistency. 

A different representation, useful, for example, for parametrizing CDF and \DZERO\ 
measurements, assumes $\Sigma \sim 1/\mpt^n$. For this parametrization (or more 
general functional forms) one can expand $\Sigma(\mpt - \mkt)$ as a power series 
in \kt\ (for \kt\ small compared to ${p}_{T}$):
\begin{equation}
\Sigma(\mpt - \mkt) = \Sigma(\mpt) + {1 \over 2!} \mkt^2 \Sigma''(\mpt)
                      + {1 \over 4!} \mkt^4 \Sigma^{(4)}(\mpt) + ...
\end{equation}
(the odd powers of \kt\ integrate out to zero). One obtains:
\begin{eqnarray}
K(\mpt) = 1 + {\mavkt^2 \over 2\pi}{n(n+1) \over \mpt^2}
      + {\mavkt^4 \over 8\pi^2} {n(n+1)(n+2)(n+3) \over \mpt^4} +...
\end{eqnarray}
For a constant (or a slowly changing) slope parameter $n$ (and for $\mavkt \ll p_T$), 
the effects of ${k}_{T}$ smearing decrease as $1/{p}_{T}^{2}$, 
as might be expected for a power-suppressed process.


\begin{references}
\bibitem{Be}
For E706, each QCD calculation has been adjusted by $A^{\alpha-1}$
to account for nuclear dependence of production  on a Be target ($A$ = 9.01), using 
$\alpha$ = 1.04 (1.08) for direct-photon ($\pi^0$) data.
\end{references}
\end{document}